\input harvmac.tex
\def\exp{{\rm exp}}

\def\frac#1#2{{#1\over#2}}

\lref\Mussar{Mussardo, G.:
Off-critical statistical models factorized scattering
theories and bootstrap program.
Phys. Rep. {\bf 218}, 215-379 (1992)}

\lref\LZ{Lukyanov, S. and Zamolodchikov, A.:
Exact expectation values of local fields
in quantum sine-Gordon model.
Preprint CLNS 96/1444, RU-96-107,
\#hepth  9611238}

\lref\kar{Karowski,  M. and Weisz, P.:
Exact form factors in (1+1)-dimensional
field theoretic models with solution behavior.
Nucl. Phys. B{\bf 139}, 455-476 (1978)}

\lref\berg{Berg, B., Karowski, M. and Weisz, P.:
Construction of Green's functions from an
exact S-matrix.
Phys. Rev. D{\bf 19}, 2477-2479 (1979)}.

\lref\yur{Yurov, V.P. and  Zamolodchikov, Al.B.:
Correlations functions of integrable 2D models of the
relativistic field theory; Ising model.
Int. J. Mod. Phys. A{\bf 6}, 3419-3440 (1991) }

\lref\FLZZ{Fateev, V., Lukyanov, S., Zamolodchikov, A. and
Zamolodchikov, Al.: Expectation values of boundary fields
in the boundary sine-Gordon model,
Preprint CLNS 97/1465, RU-97-04,
\#hepth  9702190}

\lref\KouM{Koubek, A. and Mussardo, G.:
On the operator content of the sinh-Gordon model.
Phys. Lett. {\bf B311}, 193-201 (1993)}

\lref\ZamAl{
Zamolodchikov, Al.B.:
Two-Point correlation function
in Scaling Lee-Yang model.
Nucl. Phys. {\bf B348}, 619-641 (1991)}

\lref\ZaZa{Zamolodchikov, A.B. and Zamolodchikov, Al.B.:
Factorized S-matrices in two dimensions as the exact
solutions of certain relativistic quantum field theory models.
Ann. Phys. (N.Y.) {\bf 120}, 253-291 (1979) }

\lref\Luk{Lukyanov, S.: Free Field Representation for Massive
Integrable Models,
Commun. Math. Phys. {\bf 167}, 1, 183-226 (1995)}

\lref\Fedya{Smirnov, F.A.: Form-factors in completely
integrable models of
quantum field theory. Singapore: World Scientific 1992}

\lref\MiwJ{Sato, M., Miwa, T. and Jimbo, M.: Holonomic Quantum
Fields. IV. Publ. Rims, Kyoto Univ. {\bf 15}, 871-972 (1979)}

\lref\BerLec{Bernard, D. and Leclair, A.: Differential
equations for sine-Gordon correlation functions
at the free fermion point. Nucl. Phys. {\bf B426},
543-558 (1994)}

\lref\LUkyan{Lukyanov, S.: A note on the deformed Virasoro
algebra. Phys. Lett. {\bf 367}, 121-125 (1996)}

\lref\lp{Lukyanov, S.  and   Pugai, Ya.: Bosonization
of ZF algebras: Direction toward deformed Virasoro
algebra. JETP {\bf 82}, 1021-1045 (1996)}

\lref\fre{Frenkel, E. and Reshetikhin, N.:
Quantum affine algebras
and deformations of the Virasoro
and\ $W$-algebras. Commun. Math.
Phys. {\bf 178}, 237-266 (1996)}

\lref\yap{
Shiraishi, J., Kubo, H., Awata, H.
and Odake, S.:
A quantum deformation of the
Virasoro algebra and the
Macdonald symmetric functions. Lett. Math. Phys. {\bf 38},
33-51 (1996)}

\lref\xorosh{Khoroshkin, S., Lebedev, D. and Pakuliak, S.:
Elliptic algebra\ $A_{q,p}(\widehat{sl_2})$\ 
in the scaling limit. Preprint ITEP-TH-51/96 (1996),
\#q-alg 9702002}

\lref\yapjj{Jimbo, M., Konno, H. and Miwa T.: Massless
$XXZ$ model and degeneration of the elliptic algebra
$A_{q,p}(\widehat{sl_2})$. Preprint (1996), \#hep-th 9610079 }

\Title{\vbox{\baselineskip12pt\hbox{CLNS 97/1471}
\hbox{hep-th/9703190}}}
{\vbox{\centerline{
Form-factors of exponential fields}
\vskip6pt
\centerline{in the sine-Gordon model}}}

\centerline{Sergei Lukyanov}
\centerline{Newman Laboratory, Cornell University}
\centerline{ Ithaca, NY 14853-5001, USA}
\centerline{and}
\centerline{L.D. Landau Institute for Theoretical Physics,}
\centerline{Chernogolovka, 142432, RUSSIA}
\centerline{}
\centerline{}
\centerline{}

\centerline{\bf Abstract}
An integral representation  for
form-factors of  exponential 
fields\ $ e^{ia\varphi}$\ in the
sine-Gordon model is proposed.

\Date{March, 97}
\vfill

\eject
In this letter we study the the sine-Gordon model
defined by the Euclidean  action
\eqn\hasy{
{\cal A}_{SG} =
\int d^2x\,
\bigg\{{1\over {16\pi}}\,
\big(\partial_{\nu}\varphi\big)^2 
-2\mu\,\cos\big(\beta\varphi\big)\bigg\}\ . }
Recently\ \LZ, \FLZZ,\ expectation values for
exponential fields in this
Quantum Field Theory (QFT)
were proposed,
\eqn\mainasu{\eqalign{
\langle\,  e^{ia\varphi}\, \rangle
\equiv&\ {\cal G}_a=
\bigg[\,
{M\sqrt{\pi}\, \Gamma\big({1\over 2-2\beta^2}\big)
\over 2\, \Gamma\big({\beta^2\over 2-2\beta^2}\big)}
\, \bigg]^{2a^2}
\times\cr
&\exp\biggl\lbrace\int_{0}^{\infty}{{dt}\over t}
\bigg[\ {{\sinh^2 ( 2a\beta t )}
\over{2\sinh(\beta^2 t)\, \sinh(t)\,
\cosh\big((1-\beta^2)t\big)}}-
2a^2\,e^{-2t}\ \bigg]\, \biggl\rbrace\ ,}}
where $M$ is the soliton mass.
In writing\ \mainasu\ it was assumed that the exponential
field
is normalized in accordance with the short
distance limiting form of the two-point function
\eqn\dmnf{
\langle\,  e^{ia\varphi}(x)\,e^{-ia\varphi}(y)\, \rangle \to
|x-y|^{-2a^2} \qquad {\rm as} \qquad |x-y| \to 0\ ,}
so that the field $e^{ia\varphi}(x)$ has the dimension
$[\, mass\, ]^{2 a^2}$.
The result\ \mainasu\ is
expected to hold in the domain
\eqn\djdhfg{\beta^2 < 1\ ,}
where the discrete symmetry $\varphi \to \varphi + 2\pi
n\beta^{-1}\ \ ( n=\pm1, \pm2...)$ of\ \hasy\ is spontaneously
broken (and
the QFT is massive) and\ $\langle
... \rangle$ in\ \mainasu\ means the
expectation value over one of the ground
states\ $|0\,  \rangle$\
in which the field $\varphi(x)$ is localized near 0.
The expectation value\ \mainasu\
controls both short and long
distance asymptotics of the two-point correlation function
\eqn\twopt{
{\cal G}_{a a'}\big(|x-y|\big)=
\langle\, e^{ia\varphi}(x)\, e^{ia'\varphi}(y)\, \rangle }
with $|a+a'|<\beta/2$. Indeed, if this
inequality is satisfied the short
distance limit of\ \twopt\ is dominated by OPE
\eqn\ope{e^{ia\varphi}(x)\, e^{ia'\varphi}(y)\to
\big|x-y\big|^{4aa'}\ e^{i(a+a')\varphi}(y)
\qquad {\rm as} \qquad |x-y|\to 0\  .}
Therefore
\eqn\twoass{
{\cal G}_{aa'}(r)\quad\to\quad
\Biggl\lbrace{{{\cal G}_{a+a'}\ r^{4aa'}
\qquad {\rm as} \qquad r\to 0}
\atop{\ \ \, {\cal G}_{a}\ {\cal G}_{a'}\qquad\qquad
{\rm as} \qquad r\to
\infty}}\ .}
A systematic technique for
analysis
of the  short distance expansion of\ \twopt\ is the
Conformal Perturbation Theory\ \ZamAl. At the same time
the most efficient way  to study the large\ $r$\ behavior 
is provided by  the  form-factor
approach\ \kar, \berg, \Fedya, \yur, \Mussar.
Up to now, the   whole form-factor sets (for
general values of
$\beta^2$\ from\ \djdhfg)\ have been obtained
for  fields\ $\varphi,\  e^{\pm i{\beta\varphi\over 2}}\ 
{\rm and }\  e^{\pm i \beta \varphi}$\ only\ \kar,\ \Fedya.
In this this paper we   present
form-factors of the exponential operators for  
arbitrary values of the parameters\ $0<\beta^2<1 $\ and $a$
\foot{The restriction on\ $ a$\ is not
completely clear at the moment.}.

The technique, 
providing
an integral
representation  for form-factors
in the sine-Gordon model, 
was developed in\ \Luk\foot{See also 
works\ \yapjj, \xorosh\ on an interesting 
algebraic interpretation of this procedure.}. For
the exponential  operators it is
suggested that the  matrix elements
$$\langle\,  0\,  |\,  e^{ia\varphi}\,  
|\, A_{\sigma_{2n}}( \theta_{2n})
... A_{\sigma_{1}}( \theta_1)\,
 \rangle\ ,\ \ 
\sigma_k=\pm\ \  \Big(\, \sum_{k=1}^{2n}\sigma_k=0\, \Big)\ , $$ 
where
$|\, A_{\sigma_{2n}}(\theta_{2n})  ...  A_{\sigma_{1}}
(\theta_{1})\,
\rangle $\ are
multi-soliton states
\foot{Our convention on the normalization of the states is
\ $\langle\, A^{\sigma}(\theta)\, |\, A_{\sigma'}(\theta')\,
 \rangle=2\pi\,
\delta^{\sigma}_{\sigma'}\, \delta(\theta-\theta') $.},
are given by the following  bosonization procedure
\eqn\ajahagf{
\langle\, 0\,
|\,e^{i a\varphi}\, |\, A_{\sigma_{2n}}(\theta_{2n})...
A_{\sigma_1}(\theta_1)\,
\rangle={\cal G}_a\
\langle \langle\, Z_{\sigma_{2n}}
(\theta_{2n})...Z_{\sigma_1}(\theta_1)
\, \rangle \rangle\ ,}
with
\eqn\sjshdfg{\eqalign{
&Z_{+}(\theta)=\sqrt{i {{\cal C}_2\over
4 {\cal C}_1}}\ 
e^{{a\theta\over\beta}}\
e^{i\phi(\theta)}\ ,\cr
&Z_{-}(\theta)=\sqrt{i {{\cal C}_2\over
4 {\cal C}_1}}\
e^{-{a\theta\over\beta}}\,
\biggl\{\, e^{{i\pi\over 2\beta^2}}\,
\int_{C_{+}} {d\gamma\over 2\pi}\
e^{(1-{2a\over\beta}-{1\over\beta^2})
(\gamma-\theta)}\
e^{-i{\bar \phi}(\gamma)}\,
e^{i\phi(\theta)}\ -\cr &
\ \ \ \ \ \ \ \ \ \ \ \ \ \ \ \ \ \ \ \ \ \ \ \ \ \ \  \
\  \ \ \  \ \ \ \ \ \ \ \ \ \ \ \  e^{-{i\pi\over 2\beta^2}}\,
\int_{C_{-}} {d\gamma\over 2\pi}\
e^{(1-{2a\over\beta}-{1\over\beta^2})
(\gamma-\theta)}\
e^{i\phi(\theta)}\, e^{-i{\bar \phi}(\gamma)}\,  \biggr\}\ . }}
The averaging\ $\langle \langle\, ...\, \rangle \rangle$\ should
be  performed by  Wick's theorem using the prescriptions
\eqn\ddgdfasr{\eqalign{
&\langle \langle\,
e^{i\phi(\theta_2)} e^{i\phi(\theta_1)}\, \rangle \rangle
=G(\theta_1-\theta_2)\ ,\cr
&\langle \langle\, e^{i\phi(\theta_2)} e^{i{\bar \phi}
(\theta_1)}\, \rangle \rangle
=W(\theta_1-\theta_2)={1\over
G\big(\theta_1-\theta_2-i{\pi\over2}\big)
G\big(\theta_1-\theta_2+i{\pi\over2}\big)}\ ,\cr
&\langle \langle\, e^{i{\bar\phi}
(\theta_2)} e^{i{\bar \phi}(\theta_1)}\, \rangle \rangle
={\bar G}(\theta_1-\theta_2)={1\over
W\big(\theta_1-\theta_2-i{\pi\over2}\big)
W\big(\theta_1-\theta_2+i{\pi\over2}\big)}\ .}}
The functions and the constants here  read explicitly
\eqn\ajahsgf{\eqalign{&G(\theta)=
i\, {\cal C}_1 \, \sinh\big({\theta\over 2}\big)\
\exp\biggl\{\, \int_{0}^{\infty}
{dt\over t} \ {\sinh^2 t\big(1-{i\theta\over\pi}\big)
\, \sinh(t (\xi-1))\over \sinh(2 t) \, \cosh(t)\,  \sinh(t \xi)}
\, \biggr\}\ ,\cr
&W(\theta)=-{2\over \cosh(\theta)}\
\exp\biggl\{\,-2 \int_{0}^{\infty}
{dt\over t} \ {\sinh^2 t\big(1-{i\theta\over\pi}\big)\,
\sinh(t (\xi-1))\over \sinh(2 t) \,
\sinh(t \xi)}\, \biggl\}\ ,\cr
&{\bar G}(\theta)=-{{\cal C}_2\over 4}
\ \xi\, \sinh\big({\theta+i\pi
\over \xi}\big)\, \sinh(\theta)\ ,}}
\eqn\ahsfd{\eqalign{&{\cal C}_1 =\exp\biggl\{\, -\int_{0}^{\infty}
{dt\over t} \ {\sinh^2\big({t\over 2}\big)\,
\sinh(t (\xi-1))
\over \sinh(2 t) \, \cosh(t)\,  \sinh(t \xi)}\
\biggr\}=G(-i\pi)\ ,\cr
&{\cal C}_2 =\exp\biggl\{4\int_{0}^{\infty}
{dt\over t} \ {\sinh^2\big({t\over 2}\big)\, \sinh(t (\xi-1))
\over \sinh(2 t) \,   \sinh(t \xi)}\
\biggr\}={4\over \Big[W\big(i{\pi\over 2}\big)
\xi \sin\big({\pi\over \xi}
\big)\Big]^2}\ ,}}
where
$$\xi={\beta^2\over 1-\beta^2}\ .$$
In\ \sjshdfg\ we suggest that 
the integration contour \ $C_{+}(C_{-})$\
goes from \ $\gamma=-\infty$ to $\gamma=+\infty$
and the pole $\gamma=\theta+i{\pi\over 2}$\
($\gamma=\theta-i{\pi\over 2}$) 
lies below (above) the contour.

The integrals in\ \sjshdfg\ are worked out
explicitly
for the free fermion point $\beta^2={1\over 2}$. In this case
\eqn\dqakkajk{
\eqalign{
&Z_{+}(\theta)=2^{-1}\,
e^{{i\pi \over 4}}\
e^{\sqrt{2}a\theta}\ e^{i\phi(\theta)}\ ,\cr
&Z_{-}(\theta)=-e^{-{i\pi \over 4}}\, e^{-\sqrt{2}a\theta}
\ \biggl\{\, e^{-i\sqrt{2}\pi a } e^{-i\phi(\theta+i\pi)}+
e^{i \sqrt{2}\pi a }\,  e^{-i\phi(\theta-i\pi)}\, \biggr\} }}
and
\eqn\hsgh{
\langle \langle\, e^{i\phi(\theta_2)}
e^{i\phi(\theta_1)}\, \rangle \rangle
=i \sinh\big({\theta_1-\theta_2\over 2}\big)\ .}
With the formulas\ \dqakkajk, \hsgh\
at hands it is easy to get the following  simple
expression
\eqn\agfsd{\eqalign{
\langle\, 0\,
|\,e^{i a\varphi}&\, |\, A_{+}(\theta_{2n})...A_{+}(\theta_{n+1})
A_{-}(\theta_n)...A_{-}(\theta_1)\,
\rangle=
{\cal G}_a\
(-1)^{{n(n+1)\over 2}}\  \Big( i
\sin\big(\sqrt 2 \pi a\big)\Big)^n\times\cr &
e^{\sqrt{2} a \sum_{k=1}^n(\theta_{n+k}-\theta_{k})}\
{\prod_{1\leq k<j\leq n}
\sinh\big({\theta_k-\theta_j\over 2}\big)\,
\sinh\big({\theta_{n+k}-\theta_{n+j}\over 2}\big)\over
\prod_{1\leq k, j\leq n}
\cosh\big({\theta_{n+k}-\theta_j\over 2}\big)}\ ,}}
which  is  in   agreement with the  results of\ \MiwJ, \BerLec.

It is instructive to consider two-point form-factors 
more closely.
{}From \ajahagf\  we have
the  integral representation
\eqn\sksjdyy{\langle\, 0\, |\,e^{i a\varphi}\,|\, 
A_{\pm}(\theta_{2})A_{\mp}(\theta_1)\,
\rangle={\cal G}_a\ F^{a}_{\mp \pm} (\theta_1-\theta_2)\ ,}
with
\eqn\jshsgftrq{\eqalign{&F^{a}_{-+}(\theta)=
-{G(\theta)\over G(-i\pi)}\
e^{{\theta+i\pi\over 2\xi}}\
\biggl\{\,  e^{{2i\pi a\over \beta}}\
I_{a}(\theta)+I_{a}(-2i\pi-\theta)\, \biggr\}\ ,\cr
&F^{a}_{+-}(\theta)=-{G(\theta)\over G(-i\pi)}\
e^{-{\theta+i\pi\over 2\xi}}\
\biggl\{\,  e^{-{2i\pi a\over \beta}}\
I_{a}(\theta)+I_{a}(-2i\pi-\theta)\, \biggr\}\ .}}
Here the argument of\ $I_{a}(-2i\pi-\theta)$\ means
the corresponding analytical continuation
of the function\ $I_{a}(\theta)$. The latter  is specified
for real\ $\theta$\ and
$$-\beta^{-1}+{\beta\over 2}<\Re e\, a<{\beta \over 2}$$
by
the integral
\eqn\shsgdd{I_a(\theta)=
{1\over \Big[W\big(i{\pi\over 2}\big) 
\xi \sin\big({\pi\over \xi}
\big)\Big]^2}\
\int_{-\infty}^{\infty} {d \gamma\over 2 \pi} \
W\big(\gamma+{\theta\over 2}-i\pi\big)
W\big(\gamma-{\theta\over 2}-i\pi\big)\
e^{(1-{2a\over\beta}-{1\over\beta^2})\gamma}\ .}
Eqs. \jshsgftrq\ were  checked against
the known  form-factors\ \kar, \Fedya.
In particular, if\ $a\to 0$\
\eqn\sksjsugh{F^{a}_{\mp\pm}(\theta)=
\pm i a\  F(\theta)+
O\big( a^2\big)\, , }
and\ $F(\theta)$\ coincides with the   two-particle
form-factor of the sine-Gordon field \ $\varphi$.
It can be evaluated by the method
discussed in the Appendix 4  of\ \Luk\ with the result,
\eqn\sjshst{F(\theta)=
-{G(\theta)\over
G(-i\pi)}\
{\pi \over \beta \,
\cosh\big({\theta+i\pi\over 2\xi}\big)\, \cosh\big({\theta
\over 2}\big)}\ .}
Similarly, one can  find 
\eqn\sshgatt{F^{{\beta\over 2}}_{\mp \pm} (\theta)=
{G(\theta)\over G(-i\pi)}\
{2 i\ e^{\mp{\theta+i\pi\over 2\xi}} \over
\xi  \sinh\big({\theta+i\pi\over \xi}\big)}\ }
and
\eqn\slsoi{F^{\beta}_{\mp \pm} (\theta)=
{G(\theta)\over G(-i\pi)}\ \cot\big({\pi\xi\over 2}\big)\
{4 i \cosh\big({\theta\over 2}\big)
\ e^{\mp{\theta+i\pi\over 2\xi}} \over
\xi  \sinh\big({\theta+i\pi\over  \xi}\big)}\ .}
Eqs.\ \sjshst,\ \sshgatt,\ \slsoi\ agree
with known expressions from\ \kar, \Fedya.
Notice
that the functions\ \jshsgftrq\ admit 
simpler forms
for any integer and half-integer values of 
\ ${a\over \beta}$. Under this condition,  one can  show that
\eqn\shsgdr{\eqalign{F^{l\beta}_{\mp \pm} (\theta)=
{G(\theta)\over G(-i\pi)} {4 i \ e^{\mp{\theta+i\pi\over 2\xi}}
\over \xi  \sinh\big({\theta+i\pi\over  \xi}\big)}\
&\sum_{m=1}^{l}\ (-1)^{l-m} \cosh\big((m-{1\over 2})\theta\big)
\times\cr
&\prod_{k=m-l, k\not=0}^{m+l-1}
\cot\big({\pi\xi k\over 2}\big)\ ,}}
\eqn\gdraaa{\eqalign{
F^{(l+{1\over 2})\beta}_{\mp \pm} (\theta)=&
{G(\theta)\over G(-i\pi)} {2 i \ e^{\mp{\theta+i\pi\over 2\xi}}
\over \xi  \sinh\big({\theta+i\pi\over  \xi}\big)}\
\biggl\{\, \prod_{k=1}^{l}
\cot^2\big({\pi\xi k\over 2}\big)+\cr
&2\sum_{m=1}^{l}\ (-1)^{l-m} \cosh( m\theta )
\, \prod_{k=m-l, k\not=0}^{m+l}
\cot\big({\pi\xi k\over 2}\big)\, \biggr\}\ ,}}
for $l=1,2,...$ and
$$\eqalign{
&F^{-l\beta}_{\mp \pm} (\theta)=F^{l\beta}_{ \pm\mp} (\theta)\ ,
\cr
&F^{-(l+{1\over 2})\beta}_{\mp \pm} (\theta)=
F^{(l+{1\over 2})\beta}_{ \pm\mp} (\theta)\ .} $$
Significant simplifications also  appear 
at  the so-called
reflectionless points,
$$\beta^2={n\over n+1}\ ,\ \ \  n=1,2,...\ ,$$
for generic\ $a$. Then,
\eqn\sjshsyt{F^{a}_{\mp\pm}(\theta)=\pm {G(\theta)\over G(-i\pi)}
\, {2 i\, n  \ e^{\mp{n(\theta+i\pi)\over 2}}
\over   \sinh\big(n (\theta+i\pi )\big)}\
\sum_{p=1}^{n} e^{\pm\theta (p-{1\over 2}-{a\over\beta})}\
\prod_{k=p-n}^{p-1} {\sinh\big({\pi\over n}
({a\over\beta}-{k\over 2})\big)\
\cos\big({\pi k\over 2 n}\big)
\over \cos^2\big({\pi (p-k-1)\over 2 n}\big)}\ .}

If\ $0<\beta^2<{1\over 2}$,\ the spectrum of the model\ \hasy\
contains
a number of soliton-antisoliton
bound states (``breathers'') $B_n$, $n = 1, 2, \ldots, < 1/\xi$.
The lightest of these bound states\ $B_1$
coincides with the
particle associated
with the field $\varphi$
in perturbative treatment
of the QFT\ \hasy. 
An existence of the\ $B_1$\ bound state means
that
the operators\ \dqakkajk\ satisfy
the requirement\ \LUkyan,\ \yapjj 
\eqn\laosiu{
Z_{+}(\theta_2)Z_{-}(\theta_1)\to
{i\, \Gamma_{+-}^{1}\, \Lambda
\big(\theta_1+i{\pi\over 2}(1-\xi)\big)\over
\theta_2-\theta_1-i\pi(1-\xi)}\ \ \  \ {\rm as}\ \ \ 
\theta_2-\theta_1-i\pi(1-\xi)\to 0\ ,}
where
$\Gamma_{+-}^{1}=\sqrt{2
\cot\big({\pi \xi\over 2}\big)}\ $\foot{
The soliton-antisoliton amplitude\ \ZaZa\
contains a simple pole in the physical strip
corresponding to the\ $B_1$\ particle with the residue
$ S_{+-}^{+-}(\theta)\to -{ i\    (\Gamma_{+-}^{1})^2\over
\theta-i\pi(1-\xi)}\, ,\   $  \ $\theta-i\pi(1-\xi)\to 0.$}
and
\foot{The operator\ $\Lambda(\theta)$\ is closely related
with a current of  the  "deformed Virasoro algebra"
\ \lp,\ \fre,\ \yap.}
\eqn\jshsaii{\Lambda(\theta)=
{  \lambda \over 2  \sin(\pi \xi)}\
\biggl\{\
e^{-i{a\pi\xi\over \beta}}\ e^{-i\omega(\theta+i{\pi\over 2})}
-e^{ i{a\pi\xi\over \beta}}\
e^{i\omega(\theta-i{\pi\over 2})}\ \biggr\}\  , }
with
$$\omega(\theta)=\phi\big(\theta+i{\pi\xi\over 2}\big)-
\phi\big(\theta-i{\pi\xi\over 2}\big)\ ,$$
$$\lambda=
2\cos{\pi\xi\over2}\sqrt{2\sin{\pi\xi\over2}}\
\exp\biggl\{\, -\int_{0}^{\pi\xi} {d t\over 2\pi}\
{t\over \sin(t)}\, \,\biggr\}\ .  $$
Clearly 
$B_1$\ form-factors
of the exponential field
admit the  representation
\eqn\ksjoooo{\langle\, 0\,
  |\,e^{ia\varphi}\, |\, 
B_1(\theta_n)...B_1(\theta_1)\, \rangle={\cal G}_a\
\langle \langle\,\Lambda(\theta_n)...\Lambda(\theta_1)
\, \rangle \rangle\ .}
The averaging\ 
$\langle \langle\, ...\, \rangle \rangle$\ here is
performed
by  Wick's theorem  and
\eqn\dmdhr{
\eqalign{
&\langle \langle\, e^{i\omega(\theta)}\, \rangle \rangle=1\ ,\cr
&\langle \langle\, e^{i\omega(\theta_2)} e^{i\omega(\theta_1)}
\, \rangle \rangle
=R(\theta_1-\theta_2)\ ,}}
where the function\ $R(\theta)$\ for\ $ -2\pi+\pi\xi<\Im m\,
\theta<-\pi \xi$\ is  given by the integral 
\eqn\slffojks{R(\theta)={\cal N}\ \exp\biggl\{ 8\int_{0}^{\infty}
{dt\over t} \ {\sinh(t)\, \sinh\big(t\xi \big) \,
\sinh\big(t (1+\xi \big)\over \sinh^2(2 t)}\
\sinh^2 t\big(1-{i\theta\over\pi}\big)\biggr\}\ ,}
\eqn\slsksiw{{\cal N}=\exp\biggl\{ 4\int_{0}^{\infty}
{dt\over t} \ {\sinh(t)\, \sinh(t\xi) \,
\sinh\big(t(1+ \xi)\big)\over \sinh^2(2 t)}\ \biggr\}\ .}
Notice the useful relations 
\eqn\slskshjo{ R(\theta)R(\theta\pm i\pi)={\sinh(\theta)\over
\sinh(\theta)\mp i \sin(\pi \xi)}\ .}
Using\ \ksjoooo-\slskshjo, one can easily derive the  
first\ $B_1$ form-factors of the exponential fields
\eqn\sksjsuy{\eqalign{
&\langle\, 0\,|\, e^{ia\varphi}\,|\,  B_1(\theta)\,
\rangle
=-i\, {\cal G}_a\ \lambda\,  [a]\   ,\cr
&\langle\, 0\,|\, e^{ia\varphi}\, |\, B_1(\theta_2)B_1(\theta_1)\,
\rangle
=- {\cal G}_a\ \lambda^2\,   [a]^2\  R(\theta_1-
\theta_2)\ ,\cr
&\langle\, 0\,|\,
e^{ia\varphi}\, |\, B_1(\theta_3)B_1(\theta_2)B_1(\theta_1)\,
\rangle=i\, {\cal G}_a\
\lambda^3\,  [a]\
\prod_{1\leq k<j\leq 3}
R(\theta_k-\theta_j)
\times\cr
&\ \ \ \ \ \ \ \ \ \ \ \ \ \ \ \ \ \ \ \ \ \ \ \ \ \ \
\biggl\{ [a]^2+
{x_1x_2x_3\over (x_1+x_2)(x_2+x_3)(x_1+x_3)}\biggr\}\ ,\cr
&\langle\, 0\,|\, e^{ia\varphi}\, 
|\, B_1(\theta_4)B_1(\theta_3)B_1(\theta_2)B_1(\theta_1)\,
\rangle={\cal G}_a\
\lambda^4\,  [a]^2\
\prod_{1\leq k<j\leq 4}
R(\theta_k-\theta_j)
\times\cr
&\biggl\{ [a]^2+
{ (x_1+x_2+x_3+x_4)^2 x_1 x_2 x_3 x_4 +
(x_1x_2x_3+ x_1x_2x_4+x_1x_3x_4+x_2x_3x_4)^2
\over (x_1+x_2)(x_1+x_3)(x_1+x_4)
(x_2+x_3)(x_2+x_4)(x_3+x_4)} \biggr\}
\ ,}}
here \ $x_k=e^{\theta_k}\, (k=1,2,3,4 )$ and
\ $[a]={\sin({\pi \xi a\over \beta})
\over\sin(\pi \xi)} .$
The $B_1$ breather form-factors admit (for
fixed $a$) a power
series expansion 
in $\beta$ with finite radius of convergence.
Therefore it is natural to assume that the form-factors
of exponential fields  
in the sinh-Gordon model can be obtained
from the expression in\ \ksjoooo \ 
by the  continuation $\beta \to
ib$. A simple check shows that  the analytical
continuation of\ \sksjsuy\ leads to
the  form-factors of the field $e^{a\varphi}$ in the
sinh-Gordon model proposed in\ \KouM.

Finally, we note that form-factors of the
heavier breathers follow 
from\ \ksjoooo\ by the
well known
bootstrap procedure\ \ZaZa,\ \Fedya,\ \Mussar.

\centerline{}

\centerline{\bf Acknowledgments}

I am
grateful
A. Zamolodchikov and
Al. Zamolodchikov for
interesting discussions.
This research is supported in part by NSF grant. 
\hskip0.5cm

\listrefs

\end